# INFLUENCE OF SURFACTANT CRYSTALLIZATION ON THE AGGREGATION RATE OF DODECANE-IN-WATER NANOEMULSIONS. CHANGES IN THE VISCOSITY OF THE EXTERNAL PHASE

Daniela Díaz [1], Kareem Rahn-Chique [2], Neyda García-Valera [2], German Urbina-Villalba [2,*]

[1] Instituto Universitario de Tecnología "Dr. Federico Rivero Palacio", Carretera Panamericana, Km. 8, Caracas, Venezuela.

[2] Instituto Venezolano de Investigaciones Científicas (IVIC), Centro de Estudios Interdisciplinarios de la Física (CEIF), Carretera Panamericana Km. 11, Aptdo. 20632, Caracas, Venezuela. Email: german.urbina@gmail.com

**Abstract**

Rates of aggregation of dodecane-in-water (d/w) nanoemulsions stabilized with 7.5 mM sodium dodecylsulfate (SDS) are evaluated beyond 500 mM NaCl. As in the case of hexadecane-in-water (h/w) emulsions, it is found that flocculation rates ($k_{FC}$) apparently decrease with the ionic strength, departing appreciably from theoretical predictions. Since many-particle simulations proved to be very time consuming, an accurate knowledge of experimental parameters is necessary prior to the realization of further evaluations. Here, the change in the viscosity of the aqueous solution due to the surfactant phase behavior is considered as a possible cause of the referred phenomenon.

The influence of the external viscosity on the outcome of the simulations is appraised using a quotient between the actual viscosity of the surfactant solution at a given salt concentration and its value in the absence of salt. For that purpose, the viscosity of 0.5 and 7.5 mM SDS solutions was measured between 300 an 900 mM NaCl for temperatures of 20 and 25 °C. It is concluded that the augment of the aqueous viscosity due to the formation of surfactant crystals diminish significantly the aggregation rate. Yet this decline seems insufficient to justify the observed reduction of $k_{FC}$ at 25 C. Nevertheless, it is noteworthy that a 5-degree change in the temperature of the surfactant solution causes a remarkable decrease of $k_{FC}$.

Additionally, a set of two-particle simulations is used here to illustrate the limitations of this methodology for the appraisal of the viscosity contribution. It is confirmed that such approach is only convenient to study the effect of the interaction potential on the aggregation rate, as it was formerly conceived by Fuchs [1936].

**Keywords**   Emulsion, Viscosity, Dodecylsulfate, Crystals, Krafft, Drops, Stability, Aggregation.

## 1. INTRODUCTION

The phase-behavior of ionic surfactants results from the distinct physical affinities of its chemical structure. In aqueous solutions, surfactants aggregate into micelles when a minimum concentration known as CMC is reached (8.3 mM for sodium dodecylsulfate (SDS) [Lindman, 1980]). Micelles allow the hydrophobic tails of the amphiphile to stay out of the water in an oil-like environment while suspended in an aqueous medium. This collective behavior requires a minimum number of molecules, and therefore, it occurs after surfactant adsorption and the subsequent saturation of the available interfaces [Urbina-Villalba, 1997].

Electrolytes increase the polarity of the water phase promoting the proliferation of micelles at lower surfactant concentrations [Urbina-Villalba, 2013; 2016]. Thus, the CMC decreases with the increase of the ionic strength [Nakayama, 1967]. This phenomenon is referred in the bibliography as the "salting out" of the surfactant solution.

At low temperatures, surfactants may also crystallize and precipitate as a regular solute [Nakayama, 1967; Sammalkorpi, 2009; Iyota, 2009; Qazi, 2017]. The temperature at which the solubility of the surfactant reaches its CMC is known as the Krafft point ($T_{kr}$). It is equal to 22 °C for SDS in the absence of salt [Evans, 1999; Lindman, 1980]. Below this limit ($T < T_{kr}$), the total surfactant solubility is equivalent to its monomer concentration. Above $T_{kr}$ it is equal to the micelle concentration. However, if the surfactant concentration is increased beyond the CMC at $T < T_{kr}$, the amphiphile in excess precipitates in the form of hydrated crystals; otherwise ($T > T_{kr}$) the number and the size of the micelles increase. As a result of this behavior, the solubility curve of SDS rises pronouncedly around 25 °C. At this temperature it is possible to change between crystals and micelles varying the temperature in only one degree ($\pm1°$).

Tension measurements suggest that the surfactant surface excess is maximum at the CMC and remains constant at higher concentrations. Such concentrations should guarantee the kinetic stability of ionic emulsions towards aggregation for $T > T_{kr}$. Unfortunately, micelles are very effective at solubilizing oil, and consequently, low volume-fraction nanoemulsions spontaneously disintegrate after a few minutes [Rahn-Chique, 2017]. Conversely, at $T < T_{kr}$ hydrated crystals precipitate out of the solution. What happens with the adsorption equilibrium of the surfactant in this case? Emulsions could be stable at these temperatures if the surface excess remains constant. However, this requires an aqueous surfactant concentration similar to the CMC, a condition which does not appear to be compatible with a significant precipitation of surfactant crystals. Following this line of reasoning, emulsions are expected to be unstable for $T < T_{kr}$ due to the flocculation of their drops. Surprisingly, this does not appear to be the case.

Assuming a constant-kernel ($k_{ij} = k_{FC}$), an average flocculation/coalescence ($k_{FC}$) rate for nanoemulsions can be evaluated [Rahn-Chique, 2012a; 2012b; 2012c]. It results from fitting the temporal variation of the turbidity ($\tau$) to a suitable theoretical expression:

$$\tau = n_1\sigma_1 + x_a\sum_{k=2}^{k_{max}}n_k\sigma_{k,a} + (1-x_a)\sum_{k=2}^{k_{max}}n_k\sigma_{k,s} \quad (1)$$

Here: $\sigma_1$, $\sigma_{k,a}$ and $\sigma_{k,s}$ represent the optical cross sections of primary drops, aggregates of k primary drops, and large spherical drops with a volume equivalent to k primary drops ($R_k = \sqrt[3]{k}\,R_0$). According to Smoluchowski [1917] the number of aggregates per unit volume of each size $n_k$, is equal to:

$$n_k = \frac{n_0\,(k_{FC}\,n_0\,t)^{k-1}}{(1 + k_{FC}\,n_0\,t)^{k+1}} \quad (2)$$

Hence, fitting of Eq. (1) to experimental data provides a value for $k_{FC}$ along with the fraction of drops which does not coalesce after a collision: $x_a$.

Alternatively, the aggregation velocity can be approximated by the rate of doublet formation ($k_{11}$). This can be obtained from the initial change of the absorbance (Abs) as a function of time (t) [Lips, 1971a; 1973; Lichtenbelt, 1973]:

$$\left(\frac{d\tau_{exp}}{dt}\right)_0 = 230\left(\frac{dAbs}{dt}\right)_0 = \left(\frac{1}{2}\sigma_2 - \sigma_1\right)k_{11}\,n_0^2 \quad (3)$$

where $\sigma_1$ and $\sigma_2$ correspond to the optical cross sections of a single drop and a doublet, respectively.

In the absence of repulsive forces, suspended particles diffuse freely until they make contact. According to Stokes law, this diffusion is inversely proportional to the viscosity of the external phase. Moreover, the constant kernel solution for irreversible aggregation [Smoluchowski, 1917] yields a flocculation rate which is also inversely proportional to the viscosity ($\eta$):

$$k_s = \frac{4\,k_B\,T}{3\,\eta} \quad (4)$$

In Eq. (4), $k_s$ is the flocculation rate, $k_B$ the Boltzmann constant, $T$ the absolute temperature, and $\eta$ the viscosity of the suspending medium.

According to Derjaguin, Landau, Verwey and Overbeek (DLVO theory) [Derjaguin, 1941; 1967; Verwey, 1948], electrostatic forces create a potential energy barrier that hinders the attainment of primary minimum flocculation. The barrier must be surpassed in order to reach the absolute minimum. Hence, irreversible flocculation is delayed by the repulsive barrier providing kinetic stability to the dispersion. Debye and Hückel [Debye, 1923] showed that a high electrolyte concentration screens the surface charge of the particles. Thus, DLVO predicts that the height of the repulsive barrier between charged particles decreases with the addition of salt, promoting their aggregation.

Previous studies on hexadecane-in-water nanoemulsions showed that for 0.5 mM SDS ($20 \leq T \leq 25$ °C, $\phi = 10^{-4}$) primary minimum flocculation does not occur within the range of salinities studied $300 < [NaCl] < 1000$ mM. Hence, the size of the repulsive barrier **is not** related to the rate of flocculation in this case. Instead, the stability of the aggregates depends on the depth and shape of the secondary minimum of the interaction potential. According to our calculations flocs of any size partially disintegrate and re-aggregate along the complete flocculation process. The frequency of this process depends on the depth of the sec-ondary minimum *between each pair of drops* conforming an aggregate. As the salt concentration increases, the depth of these minima increase and the kinetic stability of the flocs augments. As a result, the total number of aggregates decreases with time, but at a much smaller rate than the one predicted by Smoluchowski for fast aggregation.

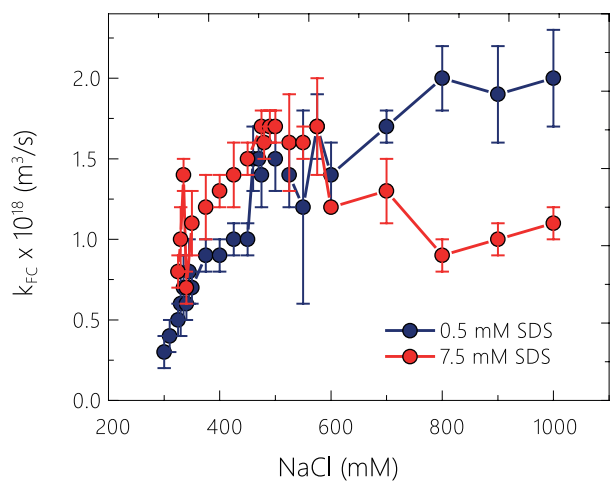

Figure 1: Change of the mixed flocculation/coalescence rate of hexadecane-in-water nanoemulsions (stabilized with 0.5 and 7.5 mM SDS) as a function of salt.

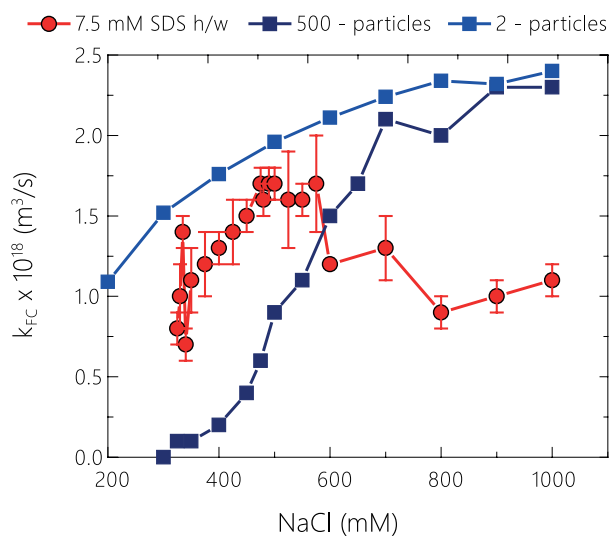

Figure 2: Experiment vs. theory for hexadecane-in-water nanoemulsions stabilized with 7.5 mM SDS. Results from 500-particle and 2-particle simulations.

If the process of aggregation is induced by the addition of salt at 0.5 mM SDS, $k_{FC}$ increases monotonically as a function of the ionic strength. This situation coincides very reasonably with the theoretical predictions of DLVO theory. However, at a *high* surfactant concentration (7.5 mM SDS) a different behavior is observed. The aggregation rate increases first reaching a plateau between 450 and 550 mM NaCl, and then decreases (Figure 1).

For 7.5 mM SDS, experimental rates are found to be substantially faster than predicted by the simulations below 450 mM NaCl, and appreciably slower than predicted above 550 mM NaCl (Figure 2). While the low-salt regime can be partially justified on the basis of micelle solubilization [Rahn-Chique, 2017], no straightforward explanation exists for the high-salt regime. Such performance cannot be accounted for either assuming deformable drops, or considering secondary minimum flocculation [Urbina-Villalba, 2015; 2016; Toro-Mendoza, 2010].

A methodology based on two-particle simulations can reproduce the order of magnitude of the experimental flocculation rates as well as their general dependence with salt in the case of h/w at 0.5 mM SDS. For 7.5 mM SDS, a maximum value around 500 mM NaCl and the significant decrease which follows, is only found assuming deformable drops (see Fig. 10 in [Urbina-Villalba, 2016]). However in this case, the rates predicted are between four and five times faster than the experiment, which is unacceptable.

Two particle simulations require the definition and implementation of a reasonable criterion for secondary-minimum flocculation. The value of each rate is calculated through a stability-ratio (W) type formula [Fuchs, 1936; McGown, 1967]:

$$k_{FC} = (t^{fast}_{ave} / t_{ave})k^{fast}_{FC} = k^{fast}_{FC} / W \qquad (5)$$

In Eq. (5), average times for secondary-minimum flocculation are required (see [Urbina-Villalba, 2016] for details). This allows the calculation of a stability ratio W, which determines the relative order of the rates from one another. The order of magnitude is provided by $k^{fast}_{FC}$ which is evaluated from one many-particle simulation run at the highest ionic strength considered ([Urbina-Villalba, 2009a, 2009b, 2016]).

As shown in Fig. 2, two-particle simulations of spherical drops predict a monotonous increase of the aggregation rate as a function of the salt concentration as expected from DLVO theory and the stability ratio approach. Moreover, they overestimate the value of the actual rates above *and below* 500 mM NaCl.

Remarkably, a concentration of 7.5 mM SDS is sufficient to produce the precipitation of crystals at high ionic strength a few minutes after mixing [Rahn-Chique, 2017]. Therefore, it is uncertain if the variation of the absorbance during shorter times (typically used for the evaluation of $k_{11}$ and $k_{FC}$ (t < 100 s)) can solely be attributed to a decline of the aggregation rate, or it comprises mixed optical contributions from drops and crystals of sub-micron size.

In the latter case, an increase in the number of scattering objects is difficult to harmonize with an absorbance reduction, unless micelles promote a considerable solubilization of the drops prior to surfactant precipitation. In the former case, it is unclear if the decrease of $k_{FC}$ can be attributed to an augment of the viscosity of the external phase, or to an upsurge of an additional repulsive interaction between the drops due to crystal formation.

Most studies related to crystallization in o/w emulsions deal with the solidification of the drops as a function of temperature [Kékicheff, 1989; Copeland, 2002; Abramov, 2016; Miller, 2018]. A considerable amount of works concerns the crystallization of proteins at the interface of aggregated drops, which induces the phenomenon of "partial coalescence" [Boode, 1991; 1992; Thivilliers, 2008; 2010; McClements, 2012]. Even the effect of "liquid" and "crystallizable" surfactants on the partial coalescence of emulsions has been studied [Thivilliers, 2008]. However, up to our knowledge, little is known about the possible effect of surfactant crystallization on the surface of oil drops, though it is reasonable to suppose that the interface of a drop might serve as a nucleation center for surfactant precipitation. In this case, instability could be induced through a change of structure of the interfacial layer. Such behavior would be consistent with the instantaneous change in the opacity of the emulsions which is observed in h/w and d/w systems stabilized with SDS after the addition of salt [Urbina-Villalba, 2017].

Back in 2013 [Urbina-Villalba, 2013], we evaluated the rate of doublet formation for a hexadecane-in-water nanoemulsion using Eq. (3). As exposed in Figure 3, a huge difference emerges between the values of $k_{11}$ corresponding to 7.5 y 0.5 mM SDS beyond an ionic strength of 500 mM NaCl. This difference could probably be justified if the viscosity of the 7.5 mM solutions at high ionic strength is substantially different than the one of 0.5 mM SDS.

In this article we revisit the dodecane-in-water system in order to appraise the aggregation rates of 7.5 mM SDS na-noemulsions beyond 500 mM NaCl. The purpose of this determination is to confirm that the decrease of the aggregation rate at high ionic strength is a general phenomenon, not circumscribed to hexadecane-in-water emulsions. Second, we evaluate the viscosity of SDS solutions at high salinities exploring the range of temperatures usually employed in the experiments (20 – 25 °C). Then, we develop an approximate experimental correction for theoretical aggregation rates based on their inverse dependence on the viscosity of the aqueous surfactant solution. Finally, the limitations of two-particle simulations for the appraisal of a viscosity phenomenon are outlined.

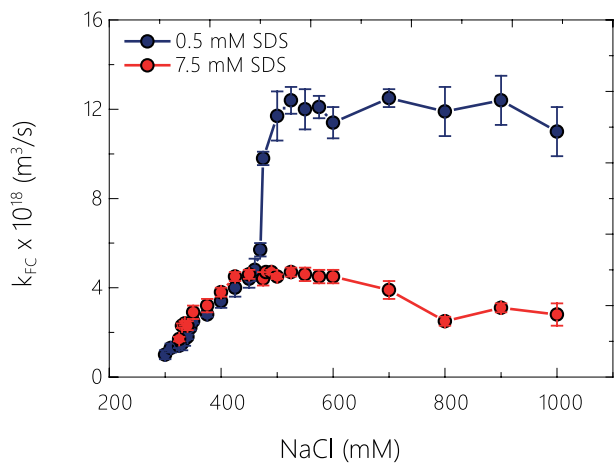

## 2. EXPERIMENTAL PROCEDURE

Dodecane (Merck, 98%) was purified using an alumina column. SDS, sodium chloride (Merck, 99.5%) and isopentanol (Scharlau Chemie, 99%) were used as received. Distilled water was deionized using a Millipore Simplicity apparatus.

Nanoemulsions were prepared using a phase inversion composition method [Rahn-Chique, 2012]. A mixture of liquid crystal solution and oil ($\phi$ = 0.84, [SDS]= 10 wt%, [NaCl]= 8% wt%, and [isopentanol]= 6,5 wt%) previously pre-equilibrated, was suddenly diluted until $\phi$ = 0.44, to obtain 120-150 nm drops of oil. This mother nanoemulsion was further diluted with pure water until $\phi$ = 0.02, and afterwards with an appropriate surfactant solution until $\phi_2$ = 3.2 x 10⁻⁴.

## 2.1 Absorbance *vs.* time

A Shimadzu UV-visible spectrophotometer was used at $\lambda$=800nm. Samples of 2.4 mL were used, along with injections of 0.6 mL with distinct NaCl solutions. The absorbance (Abs) was followed during 5 minutes for salt concentrations between 100 and 900 mM NaCl at a constant surfactant concentration of 7.5 mM SDS. Whenever possible the data of Abs vs. t was fitted to Eqs. (1) and (3) in order to obtain $k_{FC}$ and $k_{11}$ respectively. For that purpose a symbolic algebra code previously written in Mathematica (Wolfram, 8.0.1.0) was used. The value of the average radius is an input of the calculation which was slightly varied ($R_{teo}= R_{exp} \pm \delta$) in order to maximize the quality of the adjustment, and to obtain a consistent initial aggregation time.

## 2.2 Viscosity Measurements

A Falling Ball Viscometer Type C from Haake was used to appraise the viscosity of 7.5 and 0.5 mM SDS solutions at 6 salt concentrations. For this purpose, the lapse of time ($t_{fall}$) required by a falling sphere to travel between two marks (A and B on <span style="color:red">Fig. 4</span>) was measured on a thermo stated cylinder

$$\eta \;=\; K(\rho_1 - \rho_2)t_{fall} \qquad (6)$$

In Eq. (6) $\rho_1$ and $\rho_2$ stand for the density of the ball and the external phase. K is a constant for each sphere (Table 1). The temperature was kept fixed using a cryostat at either T = 20° or T = 25°C.

Standard balls No. 1 and 2 (Table 1) were used. The selected sphere depended on the time of descent. All systems were allowed to equilibrate during 15 minutes. For less viscous solutions (lower salinities) ball No. 1 ball was used. In the presence of a large amount of crystals, ball No. 2 was necessary. At T = 20 °C crystals begin to appear at 700 mM NaCl for 0.5 mM SDS and at 500 mM NaCl

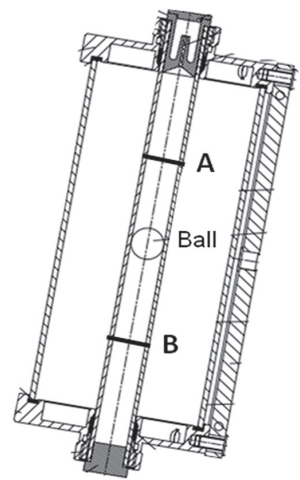

<span style="color:red">Figure 4</span>: Fluid viscometer Type C from Haake.

for 7.5 mM SDS. At T = 25 °C precipitation commences at 900 mM NaCl for 0.5 mM SDS and at 700 mM NaCl for 7.5 mM SDS. Notice that usual aggregation rate evaluations require at most 60 s. Since the amount of crystals increases with time, the value of the external viscosity determined in these experiments is expected to be significantly higher.

A volume of 40 milliliters of surfactant solution at each salt concentration was poured into the measuring cylinder of the viscometer. The corresponding ball was set in place and the hollow stopper introduced. After equilibration, the viscometer cylinder was turned upside down promoting the falling of the ball. The time required for the displacement of the ball between marks A and B (Figure 4) was determined using a digital watch. The lapse of time starts when the lower periphery of the ball touches mark A, and ends when it reaches mark B. This time of fall ($t_{fall}$) was used to calculate the viscosity of the medium using Eq. (6). The procedure was performed five times for each system. The value of density was approximated by weighting one milliliter of each solution thrice in an analytical balance. This volume was drawn and delivered employing a precision pipette.

Table 1: Physical characteristics of the balls used for Viscosity measurements.

| Ball N° | Made of | Density (g/cm³) | Diameter of ball (mm) | Constant K (approx.) mPa . s . cm³/g .s | Measuring range mPas (cP) |
|---|---|---|---|---|---|
| 1 | Boron silica glass | 2.2 | 15.81 | 0.007 | 0.2 – 2.5 |
| 2 | Boron silica glass | 2.2 | 15.66 | 0.05 | 2.0 – 20 |

## 2.3 Correction Factor for theoretical calculations

Previous computer simulations approximated the viscosity of the aqueous solution by the viscosity of water at $T = 25\ °C$: $\eta_{ref}\ (T = 25°C) = 8.9047 \times 10^{-4}\ Pa.s$. However, both the flocculation rate (Eq. (4)) and Stokes diffusion constant for a sphere of radius R are inversely proportional to the viscosity of the surrounding medium:

$$D \propto D_{Stokes} = \frac{k_B T}{6\pi\eta R} \qquad (7)$$

Consequently, the effect of the actual viscosity on the theoretical aggregation rates previously obtained can be estimated using the following formulae:

$$k_{FC}(corr) = k_{FC}\ \frac{\eta_{ref}(T)}{\eta(T, [SDS], [NaCl])}$$

$$\approx k_{FC}\ \frac{\eta(T, [SDS], [NaCl] = 0)}{\eta(T, [SDS], [NaCl])} \qquad (8)$$

Where:

$\eta(T = 25°C, [SDS] = 7.5\ mM, [NaCl] = 0) = 8.4028 \times 10^{-4}\ Pa.s$,
$\eta(T = 20°C, [SDS] = 7.5\ mM, [NaCl] = 0) = 9.6403 \times 10^{-4}\ Pa.s$
$\eta(T = 25°C, [SDS] = 0.5\ mM, [NaCl] = 0) = 8.2933 \times 10^{-4}\ Pa.s$,
$\eta(T = 20°C, [SDS] = 0.5\ mM, [NaCl] = 0) = 1.0047 \times 10^{-3}\ Pa.s$

## 2.4 Two-particle simulations

In order to illustrate the limitations of two-particle simulations for the consideration of viscosity effects, eight sets of calculation were computed. The details of the simulations can be found [Urbina-Villalba, 2016] and correspond to the case of non-deformable (spherical) droplets. In the new calculations, only the effect of three parameters were explored: temperature ($T = 20$ or $25\ °C$), time step ($\Delta t = 83, 53, 28$ and $13$ nanoseconds) and viscosity of the external phase.

The viscosity of intermediate salt concentrations was interpolated from the experimental data (Table 2) using polynomial fits from second to fourth grade:

$\eta(T, [SDS], [NaCl]) =$

$\qquad a[NaCl]^4 + b[NaCl]^3 + c[NaCl]^2 + d[NaCl]^1 + e \qquad (9)$

Table 2: Viscosity of aqueous solutions of SDS as a function of temperature and salt concentration.

| SDS (mM) | T (°C) | [NaCl] (mM) | η (mPa . s) | η / η_ref | η_ref / η |
|---|---|---|---|---|---|
| 0.5 | 20 | 0 | 1.0047 | 1.0000 | 1.0000 |
| 0.5 | 20 | 100 | 0.9514 | 0.9470 | 1.0560 |
| 0.5 | 20 | 300 | 0.9642 | 0.9596 | 1.0421 |
| 0.5 | 20 | 500 | 0.9848 | 0.9802 | 1.0202 |
| 0.5 | 20 | 700 | 1.1741 | 1.1686 | 0.8558 |
| 0.5 | 20 | 900 | 1.1836 | 1.1781 | 0.8489 |
| 0.5 | 25 | 0 | 0.8293 | 1.0000 | 1.0000 |
| 0.5 | 25 | 100 | 0.8345 | 1.0063 | 0.9938 |
| 0.5 | 25 | 300 | 0.8514 | 1.0266 | 0.9741 |
| 0.5 | 25 | 500 | 0.8746 | 1.0545 | 0.9483 |
| 0.5 | 25 | 700 | 0.9034 | 1.0893 | 0.9180 |
| 0.5 | 25 | 900 | 0.9072 | 1.0939 | 0.9141 |
| 7.5 | 20 | 0 | 0.9640 | 1.0000 | 1.0000 |
| 7.5 | 20 | 100 | 0.9588 | 0.9946 | 1.0055 |
| 7.5 | 20 | 300 | 0.9826 | 1.0193 | 0.9811 |
| 7.5 | 20 | 500 | 1.1362 | 1.1786 | 0.8485 |
| 7.5 | 20 | 700 | 1.1929 | 1.2375 | 0.8081 |
| 7.5 | 20 | 900 | 1.4245 | 1.4776 | 0.6768 |
| 7.5 | 25 | 0 | 0.8403 | 1.0000 | 1.0000 |
| 7.5 | 25 | 100 | 0.8602 | 1.0237 | 0.9769 |
| 7.5 | 25 | 300 | 0.8619 | 1.0257 | 0.9749 |
| 7.5 | 25 | 500 | 0.8869 | 1.0555 | 0.9475 |
| 7.5 | 25 | 700 | 0.9295 | 1.1061 | 0.9040 |
| 7.5 | 25 | 900 | 0.9602 | 1.1427 | 0.8751 |

Here the salt concentration is introduced in mM ($10^{-3}$ M), and the resulting viscosity comes in mPas ($10^{-3}$ Pas). Coefficients a, b, c, e, are functions of T and [SDS] (Table 3).

## 3. RESULTS

Figure 5 shows the initial variation of the absorbance of three surfactant solutions after the addition of salt. The opening of the chamber and the sudden dilution of the sample caused by the injection of brine generates a characteristic signal of W-shape. The final absorbance rapidly relaxes reaching negligible values in about 5 seconds (except in the case of 900 mM NaCl). The magnitude of the perturbation (depth and height of the W curve) depends on the procedure of injection. The larger the turbulence produced, the larger the span of the signal. The anomalous behavior shown by the 900-mM emulsion is possibly related to the rapid appearance of surfactant crystals caused by the high ionic strength.

Table 3: Coefficients of Eq. (9).

| SDS (mM) | T (°C) | a | b | c | d | e | $r^2$ |
|---|---|---|---|---|---|---|---|
| 0.5 | 20 | 0 | $-1.5915 \times 10^{-9}$ | $+2.6324 \times 10^{-6}$ | $-8.7779 \times 10^{-4}$ | 1.0106 | 0.93035 |
| 0.5 | 25 | 0 | $-2.5185 \times 10^{-10}$ | $+3.3230 \times 10^{-7}$ | $-8.9996 \times 10^{-5}$ | 0.83053 | 0.99608 |
| 7.5 | 20 | 0 | 0 | $+6.0246 \times 10^{-7}$ | $-3.7091 \times 10^{-5}$ | 0.95962 | 0.97719 |
| 7.5 | 25 | $-1.4414 \times 10^{-12}$ | $+2.5672 \times 10^{-9}$ | $-1.3043 \times 10^{-6}$ | $+2.7734 \times 10^{-4}$ | 0.84108 | 0.99826 |

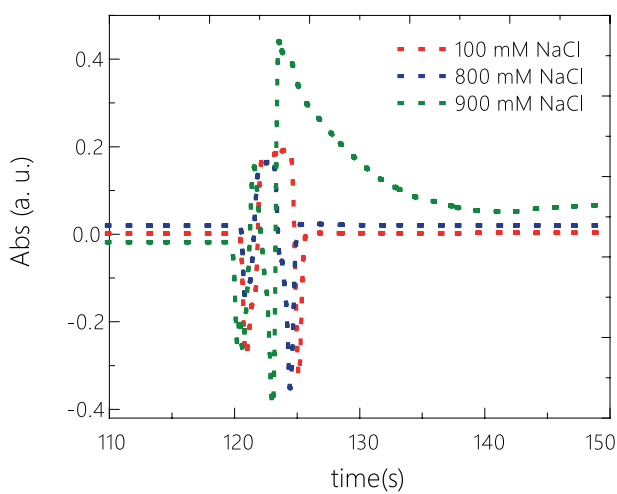

Figure 5: Evolution of the adsorbance of 7.5 mM SDS solutions after the addition of salt.

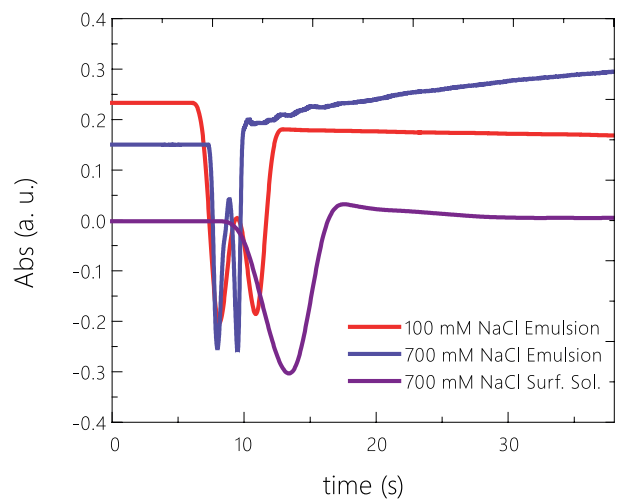

Figure 6: Comparison between the initial evolution of the adsorbance of a 7.5 mM SDS solution with the absorbance of one stable (100 mM NaCl), and one unstable (700 mM NaCl) emulsion.

Unlike solutions, the initial absorbance of nanoemulsions is substantially larger (Figure 6). The dilution perturbation has a similar shape, but the final absorbance generally reaches higher values (similar in magnitude to the initial ones). The subsequent evolution of the absorbance depends on the stability of the emulsion with respect to flocculation and solubilization. On the one hand, at [NaCl] < 300 mM the absorbance decreases smoothly towards cero (Figure 7). This decrease is caused by the solubilization of the drops [Rahn-Chique, 2017]. The sample looks completely clear after a few hours when 7.5 mM SDS is used. On the other hand, if [NaCl] > 300 mM the aggregation of the drops dominates, and the absorbance increases appreciably during several seconds (Figs. 6 and 7). In this case, when the sample vessel is withdrawn from the spectrophotometer, a thin layer of cream is observed.

The "break-even" situation occurs around [NaCl] ~ 300 mM for d/w and h/w nanoemulsions. Aggregation slightly predominates over micellar solubilization during the first 250 s, and then, solubilization takes over. The slope of the absorbance is so small that prevents the evaluation of the flocculation rate.

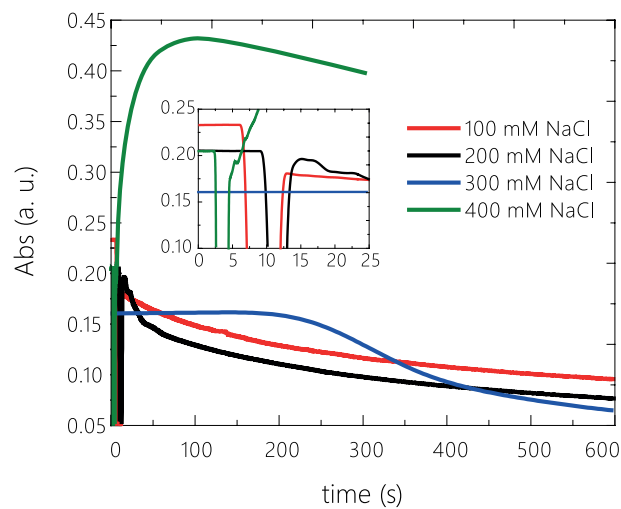

Figure 7: Curves of Absorbance vs. time corresponding to a set of dodecane-in-water nanoemulsions after the addition of salt.

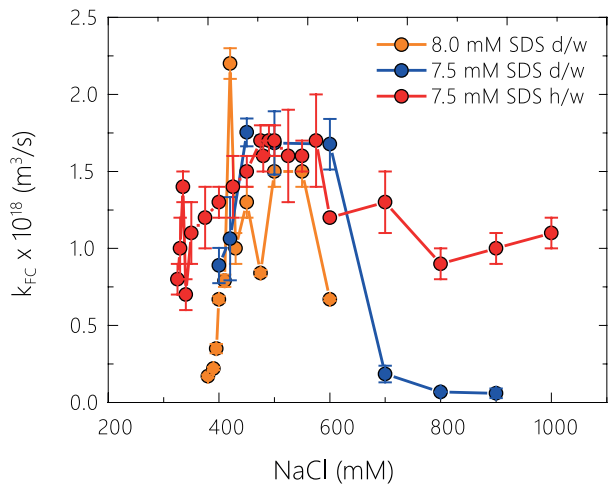

Figure 8: Experimental aggregation rates for dodecane-in-water (d/w) and hexadecane in water (h/w) nanoemulsions.

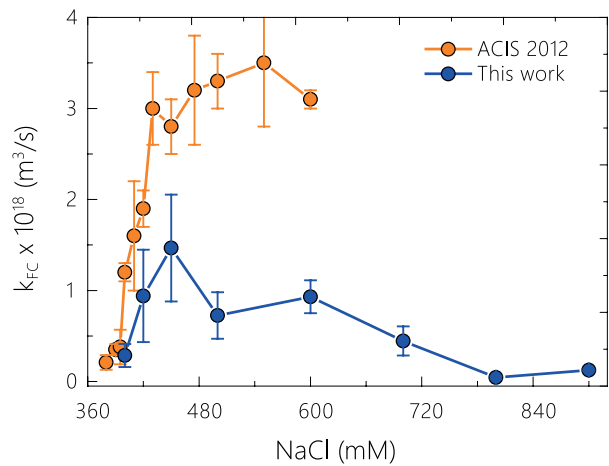

Figure 9: Initial aggregation rates for d/w emulsions at 7.5 (this work) and 8.0 mM NaCl ([Rahn-Chique,2012]).

Figure 8 contrasts the values of $k_{FC}$ obtained in this work for dodecane-in-water emulsions with our previous evaluations of this system [Rahn-Chique, 2012], and the ones obtained for hexadecane-in-water emulsions [Urbina-Villalba, 2015; 2016]. Between 450 and 550 mM NaCl, a remarkable coincidence of the data occurs. Although each symbol corresponds to the average of three measurements, a contrast between the measurements of d/w obtained at 420 and 600 mM NaCl for 7.5 and 8 mM SDS, suggests that despite the precision of the determinations, the absolute error could be as high as $1 \times 10^{-18}$ m³/s. However, it should be remarked that each set of data was obtained by a different operator using a distinct apparatus.

In any event, it is clear that the aggregation rate of the emulsions decreases appreciably beyond 600 mM NaCl (systems are apparently more stable with respect to flocculation). According to the present results, the d/w system appears to be much more sensible to this variation than the h/w one.

The rates of doublet formation ($k_{11}$) corresponding to d/w confirm the above trend (Figure 9). However, a substantial dispersion of the data is observed. Comparison of Figs. 8 and 9 outlines the advantage of using Eq. (1) instead of Eq. (3) for the evaluation of a flocculation rate. The second methodology requires a reasonable (but arbitrary) draw of the initial slope dAbs/dt. The criteria previously used for the calculation of these rates (maximum regression coefficient) is different from the one actually employed. Currently, the absorbance of an emulsion sample diluted with pure water is used to approximate the absorbance of the same sample with salt *before the aggregation process starts*. The time at which the actual curve reaches the reference absorbance is regarded as $\tau = 0$ ($t_0$) for the aggregation process. Hence, the initial slope is the one of the line which exhibits the maximum regression coefficient starting from the absorbance corresponding to $\tau = t_0$. In the former case the slope can change appreciably depending on $t_0$, since the aggregation curve commences as soon as the last branch of the W-perturbation ends.

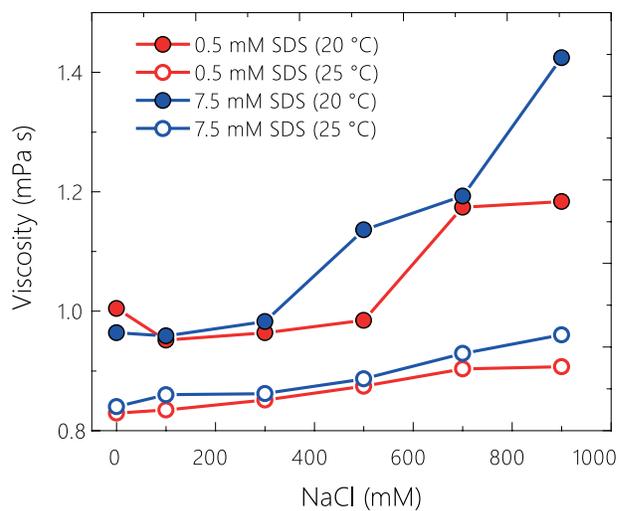

Figure 10: Change in the viscosity of surfactant solutions as a function of the salt concentration.

As shown in Figure 10 the viscosity of the surfactant solutions increases considerably with the addition of salt. Since the samples were left to equilibrate for 15 minutes, the appearance of crystals was clearly observed.

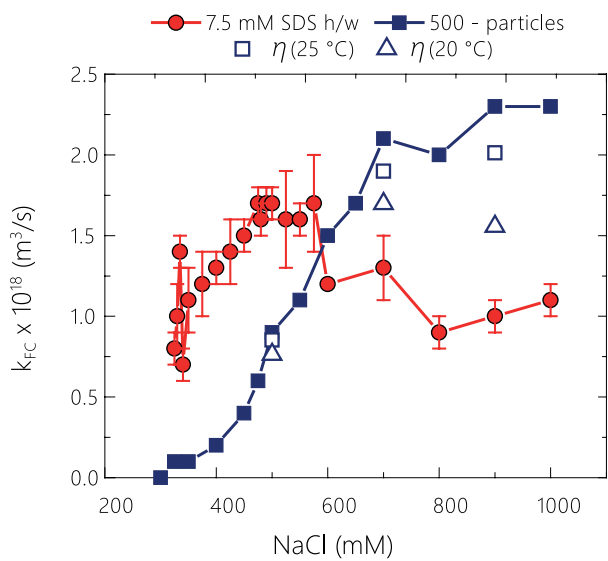

Figure 11: Viscosity corrections for 500-particle h/w simulations.

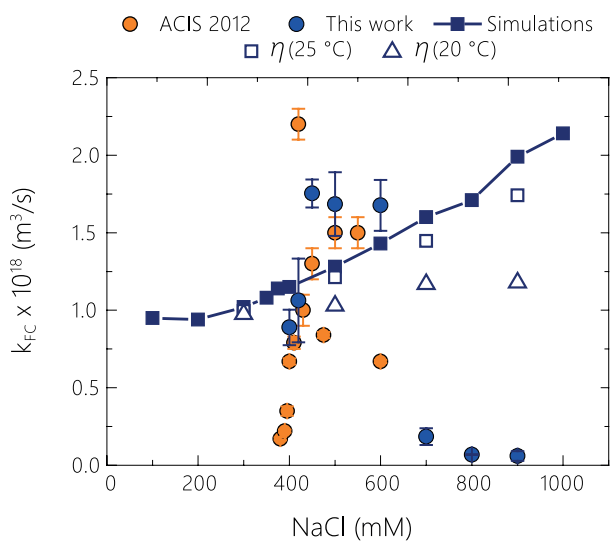

Figure 12: Viscosity corrections for previous 2-particle d/w simulations [Urbina-Villalba, 2016].

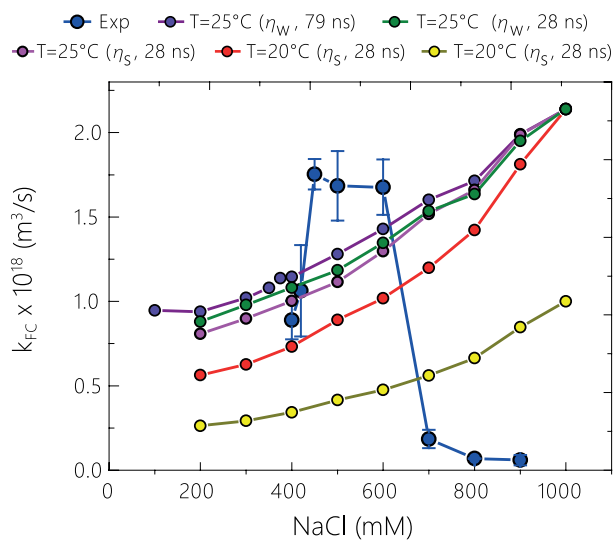

Figure 13: Two-particle simulations. Symbols $\eta_w$ and $\eta_s$ stands for the viscosity of pure water and the one of the aqueous solutions. The yellow spheres were calculated using an arbitrary value of $k_{FC}$ (fast) = $10^{-18}$ m$^3$/s.

In order to appraise the influence of the viscosity on the value of the aggregation rates predicted by the simulations, we corrected all former theoretical rates using Eq. (8). The results are shown in Figs. 11 – 12 for h/w and d/w respectively (see Table 2). The corrections favor an appreciable decrease of the rates, but do not justify the experimental values obtained.

Notice that a possible screening of the electrostatic interaction between the drops due to the presence of crystals is discarded, since it will produce faster aggregation rates as predicted by DLVO. The same result will occur is the surface excess of the surfactant decreases as a result of precipitation.

Figure 13 shows the results of two-particle simulations. The reasons why the stability-ratio approach [Rahn-Chique, 2017] is not suitable for the appraisal of viscosity effects are clearly illustrated in this figure:

a) The initial distance between the particles in the simulations is very short (7 nm in former simulations and also in this case). This length is too small to evidence a significant influence of the viscosity on the particle movement. Moreover, secondary minimum aggregation depends markedly on the Brownian contribution, which is proportional to $\eta^{-1/2}$ rather than $\eta^{-1}$. In Brownian motion the particle follows closely the potential curve, because any unbalanced impulse is quickly attenuated by the drag of the surrounding liquid.

b) There is a minor technical problem when comparing similar calculations computed with different external viscosities. As a typical molecular dynamics code, our Emulsion Stability Simulation (ESS) program uses a scaled time step, equal to:

$$\Delta t^* = \Delta t \, (D_{Stokes} / R^2) \qquad (10)$$

Where the diffusion constant $D_{Stokes}$ is given by Eq. (7). As a result, $\Delta t^*$ also depends on the viscosity. This means that the same computational conditions lead to different $\Delta t^*$ when the external viscosity changes. This limitation cannot be overcome: if the actual time step is varied on purpose so that $\Delta t^*$ is the same at each ionic strength, the effect of the relative variation of the viscosity with the salt concentration is completely nullified. The values of $\Delta t$ shown in the caption of Fig. 13 correspond to [NaCl] = 500 mM.

c) As shown in Fig. 13, a change in the time step from 79 nm to 28 nm does not affect significantly the results of the simulations. Thus, increasing the number of steps does not enhance the effect of the external viscosity. The same occurs if the viscosity of water at T = 25°C is substituted by the actual viscosity of the aqueous solution. However, if the value of the viscosity corresponding to 20°C is employed, a substantial decrease of most values of $k_{FC}$ is observed.

d) Due to the structure of Eq. (5) the highest value of $k_{FC}$ corresponds to the highest salinity studied (W = 1). As a result, the rates increase monotonically up to $k_{FC}$ ([NaCl] = 1000 mM) = $k_{FC}^{fast}$ = 2.14 x $10^{-18}$ m³/s. If the value of $k_{FC}^{fast}$ is arbitrarily changed to $1.00 \times 10^{-18}$ m³/s (for example) the yellow curve is obtained. It is clear therefore that the observed decrease of $k_{FC}$ with the ionic strength (blue curve) cannot be reproduced with this procedure.

## 4. CONCLUSIONS

The viscosity of SDS solutions depends on its inner structure, which changes with the formation of micelles and crystals. High ionic strengths lower the CMC of the solutions increasing their Krafft point [Nakayama, 1967]. At 7.5 mM SDS, NaCl concentrations higher than 450 mM induce the formation of macroscopic surfactant crystals in the aqueous phase of dilute d/w emulsions, a few minutes after mixing. Consequently the external viscosity increases significantly (Figure 10). This increase however, does not appear to justify the pronounced reduction of $k_{FC}$ vs. [NaCl] found in the experiments.

Slow aggregation at high ionic strength has been related in the past to surface charge reversal which cannot occur in this case. Alternatively, hydration forces have been forwarded as a possible explanation for the stability of protein-polymer systems [Molina-Bolívar, 1997]. In this simple system we spec-ulate that the interface of the drops could act as nucleation centers for surfactant crystallization, reducing the diffusion of the drops and possibly their coalescence frequency.